\documentclass[%
reprint,
showpacs,preprintnumbers,
nofootinbib,
 amsmath,amssymb,
 aps,
 prd,
]{revtex4-1}

\usepackage{graphicx}

\usepackage[dvips]{color}

\newcommand{\ene}[1]{E_{\mathbf{#1}}}

\begin{document}

\preprint{Imperial/TP/11/DJW/01}	
\title{Nonperturbative study of the 't Hooft-Polyakov monopole form factors}
\author{Arttu Rajantie}
\email{a.rajantie@imperial.ac.uk}
\author{David J. Weir}
\email{david.weir03@imperial.ac.uk}
\affiliation{%
Department of Physics, Imperial College London SW7 2AZ, U.K.
}

\date{September 1, 2011} 
\begin{abstract}
The mass and interactions of a quantum 't Hooft-Polyakov monopole are measured nonperturbatively using correlation functions in lattice Monte Carlo simulations. A method of measuring the form factors for interactions between the monopole and fundamental particles, such as the photon, is demonstrated. These quantities are potentially of experimental relevance in searches for magnetic monopoles.
\end{abstract}
\maketitle

\section{Introduction}
Everyday experience tells us that there are no isolated magnetic charges, i.e., magnetic monopoles, but there are strong theoretical hints that they may still exist. Their existence would explain the quantisation of electric charge~\cite{Dirac:1931kp}, and they are an inevitable consequence of grand unification~\cite{'tHooft:1974qc,Polyakov:1974ek}. 
As stable particles, magnetic monopoles produced in the early universe would still exist~\cite{Kibble:1976sj,Preskill:1979zi}, but there are very strong astrophysical bounds on their number density (as outlined by the review in Ref.~\cite{Nakamura:2010zzi}). However, monopoles could also be produced in particle accelerators provided that their mass is low enough. This would clearly not be possible for grand unified theory monopoles, for which the mass would be around $10^{16}~{\rm GeV}$. It is, however, entirely consistent to consider monopoles which are much lighter than this, perhaps even around $1~{\rm TeV}$.  They can then be produced in the LHC, where they are being searched for by the MoEDAL experiment~\cite{Pinfold:1999sp}.

If the search is successful, it would open up a new window on high energy physics. The monopoles could have interesting and unusual properties -- such as the ability to catalyse baryon decay~\cite{Rubakov:1988aq} -- which reflect  physics beyond the Standard Model and yet, because they are stable and interact strongly with the electromagnetic field, they would be relatively easy to study. Curiously, effective excitations with non-zero magnetic charge exist also in condensed matter systems~\cite{Fennell,Branford}, where their physics can be studied with simple tabletop experiments.

To benefit from any experimental discovery of magnetic monopoles and to link their properties to high energy physics, one needs a reliable way to calculate their properties from theory. Calculations of their scattering amplitudes typically treat them as point particles and are hampered by their strong magnetic coupling~(see Ref.~\cite{Milton:2006cp} for a review). However, in actual theories of high energy physics they usually appear as topological solitons known as 't~Hooft-Polyakov monopoles~\cite{'tHooft:1974qc,Polyakov:1974ek}, which are extended objects. These solutions have been studied extensively in classical field theory~\cite{Manton:2004tk}, but quantum results generally only exist in supersymmetric theories in which quantum corrections are straightforward. In non-supersymmetric theories results have been limited to leading logarithmic corrections to the monopole mass at the one-loop level~\cite{Kiselev:1988gf,Kiselev:1990fh}, although the monopole mass has also been calculated nonperturbatively using numerical lattice field theory methods~\cite{Rajantie:2005hi}.

In previous work~\cite{Rajantie:2009bk,Rajantie:2010fw} we developed a technique to calculate form factors of topological solitons using lattice Monte Carlo simulations. In this paper we apply this technique to 't~Hooft-Polyakov monopole, and calculate the form factor of the magnetic monopole for scalar and magnetic fields. The latter describes the interaction between the monopole and the photon, and it is therefore the key observable if one is ever in a position to study magnetic monopoles experimentally.
Furthermore, it needs to be calculated in the full quantum theory because the semiclassical result is that of a pointlike Dirac monopole, and therefore any non-trivial properties of the monopole appear only in quantum theory.

We investigate the SU(2) Georgi-Glashow model with the Lagrangian
\begin{multline}
\label{equ:L}
\mathcal{L} = -\frac{1}{4} \mathrm{Tr} F_{\mu\nu} F^{\mu\nu} + \mathrm{Tr}[D_\mu, \Phi][D^\mu, \Phi]  \\
- m^2 \mathrm{Tr} \, \Phi^2 - \lambda (\mathrm{Tr}\, \Phi^2 )^2
\end{multline}
with the covariant derivative $D_\mu = \partial_\mu + ig A_\mu$. The field $\Phi$ is in the adjoint representation of the $\mathrm{SU}(2)$ gauge group and can be parameterised by the Pauli matrices as $\Phi=\phi^a \sigma^a$.

In the broken phase, which occurs classically for $m^2 < 0$ in the parameterisation chosen here, a vacuum expectation value $\mathrm{Tr}\, \Phi^2 = -m^2/2\lambda = v^2$ forms, and the $\mathrm{SU}(2)$ symmetry is broken to $\mathrm{U}(1)$. In this phase, the theory has monopole solutions with an extended scalar field \cite{'tHooft:1974qc,Polyakov:1974ek}.

Even in continuum, the classical profile of this monopole solution must be obtained numerically except in the BPS limit where $\lambda\to 0$. The classical mass of the monopole can be written as
\begin{equation}
\label{equ:Mclass}
M=\frac{4\pi m_\mathrm{W}}{g^2}f(z),
\end{equation}
where $f(z)$ is a function of the ratio $z=m_\mathrm{H}/m_\mathrm{W}$, $m_\mathrm{H} = \sqrt{2}|m|$ is the Higgs mass and $m_\mathrm{W} = g|m|/\sqrt{\lambda}$ is the mass of the charged $\mathrm{W}^{\pm}$ bosons. 
For non-zero $z$, the function $f(z)$ needs to be calculated numerically or as a
Taylor expansion~\cite{Kirkman:1981ck,Forgacs:2005vx}.

Two length scales can be associated with the quantum monopole. The first is the Compton wavelength determined by the monopole mass $M$; the other is the core size of the monopole which is determined by the perturbative masses $m_\mathrm{H}$ and $m_\mathrm{W}$. 
Eq.~(\ref{equ:Mclass}) shows that at weak coupling there is a large hierarchy between these scales, $M\gg m_\mathrm{W},m_\mathrm{H}$.

\section{Lattice implementation}

In order to study the theory (\ref{equ:L}) using Monte Carlo simulations, we Wick rotate it to 4D Euclidean space and discretise it.
Our Euclidean lattice action is
\begin{multline}
\label{equ:latticeaction}
S_\text{lat} =
2\sum_{\mu}
\left[
{\rm Tr} \Phi(\vec{x})^2-
{\rm Tr} \Phi(\vec{x}) U_\mu(\vec{x}) \Phi(\vec{x}+\hat{\mu})
U_\mu^\dagger(\vec{x})\right]
\\
+\frac{2}{g^2}\sum_{\mu<\nu}\left[2-
{\rm Tr} \; U_{\mu\nu}(\vec{x})
\right]
+m^2{\rm Tr}\ \Phi^2+\lambda({\rm Tr}\ \Phi^2)^2.
\end{multline}
with link matrices parameterised as $U_\mu = 1 + i \sigma_a u_a$. We have set the lattice spacing to unity, and therefore are left with the gauge coupling $g$, bare mass $m$ and quadratic coupling $\lambda$ as free parameters. 

\begin{figure}
\includegraphics{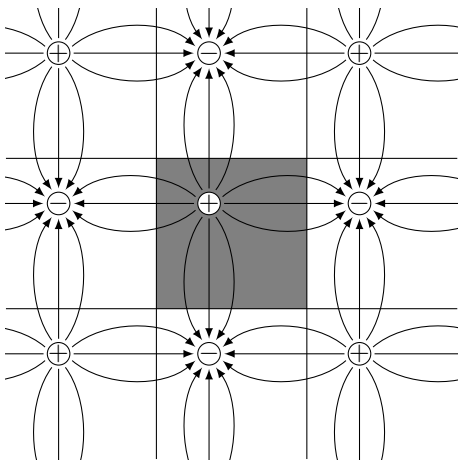}
\caption{\label{fig:bcs} A two-dimensional spatial slice through the system with boundary conditions of Eq.~(\ref{eq:bctwisted}). The magnetic field lines are shown, and the antiperiodicity of the magnetic field components can be deduced. The simulated box is shaded, with `image' magnetic charges shown in the neighbouring boxes. It can be seen that moving from the boundary on one side of the lattice to the other requires that the magnetic field be reversed.}
\end{figure}

In the symmetry broken phase, a residual $\mathrm{U}(1)$ symmetry persists. We can derive link variables $u_\mu$ corresponding to this smaller gauge group~\cite{Davis:2000kv, Edwards:2009bw}
\begin{equation}
u_\mu = \Pi_+(x) U_\mu(x) \Pi_+ (x+\hat{\mu})
\end{equation}
where $\Pi_+ = \frac{1}{2} (1 + \hat{\Phi})$ and $\hat{\Phi} = \Phi \sqrt{2/\mathrm{Tr}\,\Phi^2}$, giving an Abelian field strength tensor%
\footnote{
Other definitions for the effective U(1) field have been used in the literature~\cite{Bornyakov:1988jd}. We choose this one because 
it preserves the topological nature of the original 't~Hooft tensor~\cite{'tHooft:1974qc} and therefore defines an exactly quantised and localised magnetic charge, and because it is symmetric under lattice rotations. 
However, one should bear in mind that because we are dealing with an interacting theory, none of these expressions will be the exact creation operator for the real physical photon state. Therefore, one should really use a correlation matrix for a set of operators with the correct quantum numbers, which is the standard approach in lattice mass measurements.
}
\begin{equation}
\alpha_{\mu\nu} = \frac{2}{g} \mathrm{arg} \; \mathrm{Tr} \; u_\mu(x) u_\nu(x+\hat{\mu}) u_\mu^\dag (x+\hat{\nu}) u_\nu^\dag (x)
\end{equation}
and an expression for the lattice magnetic field,
\begin{equation}
\label{eq:latmag}
B_i = \frac{1}{2}\epsilon_{ijk} \alpha_{jk}.
\end{equation}

Gauss's law for a magnetic field in standard $\mathrm{U}(1)$ electrodynamics is $\nabla \cdot \mathbf{B} = 0$. 
In our lattice formulation, the corresponding equation becomes
\begin{equation}
\sum_{i=1}^3 \left[ B_i(\mathbf{x}+\hat{\imath}) - B_i(\mathbf{x})\right] = \rho_\mathrm{M}(x) = \frac{4 \pi n}{g},
\end{equation}
where $n$ is an integer, which can be non-zero. This means that the theory allows magnetic charges. In the classical limit, these charges correspond to 't~Hooft-Polyakov monopoles~\cite{'tHooft:1974qc,Polyakov:1974ek}. Note that even on the lattice, the magnetic charge is quantised and localised in one lattice cell.

We simulate the theory of Eq. (\ref{equ:latticeaction}) on a Euclidean lattice of size%
\footnote{\label{fn1}We choose this notation to be consistent with earlier literature. Note that our $T$ is not temperature, and even though we work in 4D Euclidean space one should not interpret it as the imaginary time formulation of finite-temperature field theory. If one were to use that interpretation, the temperature of the system would be $1/T$.}
$L^3 \times T$. 
To create nonzero magnetic charge we apply twisted boundary conditions on each timeslice~\cite{Davis:2000kv}, while retaining periodic boundary conditions in the time direction. The twisted spatial boundary conditions are
\begin{equation}
 \label{eq:bctwisted}
\begin{split}
\Phi(x+L\hat{\imath}) &= \sigma_i \Phi(x) \sigma_i \\
U(x+L\hat{\imath}) &= -\sigma_i U(x) \sigma_i
\end{split}
\end{equation}
in the $i$th direction, where $\sigma_i$ is the appropriate Pauli matrix. These boundary conditions force the magnetic charge to be odd. If $T$ is large enough then the contribution to the partition function for the simulation with these boundary conditions must come predominantly from the one-charge sector as tunnelling is heavily suppressed. Specifically, we have a partition function
\begin{equation}
\label{eq:twistpartfun}
Z_\text{tw} = 2 Z_0 \left( Z_1 e^{-M T} + \frac{1}{3!} Z_1^3 e^{-3 M T} + \ldots \right),
\end{equation}
where 
\begin{equation}
\label{equ:Z1}
Z_1 = (mL^2/2\pi T)^{3/2}
\end{equation} is the partition function for an isolated pointlike monopole (and, in fact, the usual partition function for a point particle at \textsl{temperature} $1/T$ -- see footnote~\ref{fn1}).

Similar arguments apply to the C-periodic boundary conditions~\cite{Kronfeld:1990qu}
\begin{equation}
\label{eq:cperiodic}
\begin{split}
\Phi(x+L\hat{\imath}) &= \sigma_2 \Phi(x) \sigma_2 \\
U(x+L\hat{\imath}) &= -\sigma_2 U(x) \sigma_2,
\end{split}
\end{equation}
which are locally gauge equivalent to the twisted ones~(\ref{eq:bctwisted}), but not globally.
These boundary conditions permit only even magnetic charges, with the resulting partition function taking the form
\begin{equation}
\label{eq:Cpartfun}
Z_\text{C} = Z_0 + 2 Z_0 \left( Z_1^2 \frac{1}{2!} e^{-2 M T} + \ldots \right).
\end{equation}

Note that the magnetic field defined by Eq.~(\ref{eq:latmag}) is antiperiodic with both twisted and C-periodic boundary conditions (see Figure~\ref{fig:bcs}). In contrast, with standard periodic boundary conditions the magnetic field is periodic, and therefore the total magnetic charge has to be zero.

\section{Free energy as the response \\ to a twist}
\label{sec:twist}

The conventional technique for studying magnetic monopoles and other topological defects with lattice Monte Carlo simulations has consisted of measuring their mass via their free energy~\cite{Ciria:1993yx,Kajantie:1998zn,deForcrand:2000fi,Rajantie:2005hi}.

The mass of the monopole is obtained from the difference in free energies in the two different topological sectors. This, in turn, must be obtained from the partition functions through
\begin{equation}
\label{equ:deltaF}
\Delta F = F_\mathrm{tw} - F_\mathrm{C} = -\ln \frac{Z_\mathrm{tw}}{Z_\mathrm{C}}.
\end{equation}
From this, we can obtain the mass using Eqs.~(\ref{eq:twistpartfun}) and (\ref{eq:Cpartfun}), which give
\begin{equation}
\Delta F = MT - \ln 2 - \frac{3}{2} \ln \frac{mL^2}{2\pi T} + O(e^{-2MT}).
\end{equation}
We cannot measure partition functions in Monte Carlo simulations (though we can, in principle, measure the ground state energy difference using other nonperturbative techniques~\cite{2011arXiv1103.2286H}). Instead one can integrate along a path from a point in parameter space where the free energy (and mass) of the monopole are known to vanish to the point where the mass is required,
\begin{equation}
\label{equ:Fintegral}
\Delta F= \int dg \left[ \left\langle \frac{\partial S_\text{lat}}{\partial g}\right\rangle_\mathrm{tw}  - \left\langle\frac{\partial S_\text{lat}}{\partial g}\right\rangle_\mathrm{C} \right].
\end{equation}
Derivatives of the free energy along the path are given by expectation values, which can be measured using Monte Carlo simulations. Several different integration paths have been considered in the literature.

We use this approach as a benchmark to compare our results with. We choose to vary $m^2$, and integrate from high $m^2$ where the system is in the symmetric phase and the monopole has zero mass, to low $m^2$ where the system is in the broken phase and the monopole is massive. 
In common with most of the literature, we use finite differences instead of a continuous derivative. The details of this calculation are given in Appendix~\ref{app:twist}.

\section{Two-point functions}

In Ref.~\cite{Rajantie:2010fw} we introduced an alternative approach, which uses correlation functions calculated with twisted boundary conditions, and allows us to calculate not only the mass of the monopole but also its form factors.
For any local operator $\mathcal{O}$, one can define the corresponding form factor as
\begin{equation}
f(\mathbf{p}_2,\mathbf{p}_1) = \langle \mathbf{p}_2 | \hat{\mathcal{O}}(0) | \mathbf{p}_1 \rangle,
\end{equation}
where $|\mathbf{p}\rangle$ is a quantum state with one monopole in a momentum eigenstate with momentum $\mathbf{p}$. We normalise these states in a Lorentz-invariant way as
\begin{equation}
\langle \mathbf{p}'|\mathbf{p} \rangle=(2\pi)^3 \delta^{(3)}(\mathbf{p}' - \mathbf{p}) E_\mathbf{p}.
\end{equation}
The form factor is closely related to the scattering amplitude between the monopole and the particle created by operator $\mathcal{O}$.

In this section, we start by looking at analytical and semiclassical results for the form factor of the monopole with various quantities, then go on to generalise the results of Ref.~\cite{Rajantie:2010fw} to the present case. These results will allow us to relate quantities measured in lattice simulations to scalar-monopole and photon-monopole form factors.

\subsection{Form factors: semiclassical results}
\label{sec:ffsemi}
In the semiclassical limit, the form factor is given by the Fourier transform of the classical
profile $\mathcal{O}_\text{cl}(x)$ of the quantity $\mathcal{O}$ in the monopole configuration,
\begin{eqnarray}
\label{equ:classff}
f(\mathbf{p}_2,\mathbf{p}_1) &=& \langle \mathbf{p}_2 | \hat{\mathcal{O}}(0) | \mathbf{p}_1 \rangle \nonumber\\ 
&=& \sqrt{E_{\mathbf{p}_2}E_{\mathbf{p}_1}} \int \mathrm{d}^3 x \; e^{i(\mathbf{p}_2 - \mathbf{p}_1).\mathbf{x}} \mathcal{O}_\text{cl} (\mathbf{x})\nonumber\\
&\approx& M \int \mathrm{d}^3 x \; e^{i(\mathbf{p}_2 - \mathbf{p}_1).\mathbf{x}} \mathcal{O}_\text{cl} (\mathbf{x}),
\end{eqnarray}
where the last line is valid in the non-relativistic limit, when $|\mathbf{p}_1|,|\mathbf{p}_2|\ll M$. In this case, to which we shall restrict ourselves, the form factor becomes a function of the momentum difference $\mathbf{k}\equiv\mathbf{p}_2-\mathbf{p}_1$ only,
as a direct consequence of Galilean invariance, so we will denote it by $f(\mathbf{k})$.

To determine what we should expect from our lattice simulations, let us evaluate this for the magnetic and scalar field operators. First, let us take our operator to be
\begin{equation}
\mathcal{O}=\mathrm{Tr}\,\Phi^2.
\end{equation}
There is no analytic expression for the classical (non-BPS) 't Hooft-Polyakov monopole solution,
but in continuum the scalar field has the ``hedgehog'' form
\begin{equation}
\label{equ:hedgehog}
\Phi(r)=\frac{v}{\sqrt{2}}H(r)\frac{\sigma\cdot\mathbf{x}}{r},
\end{equation}
where $r=|\mathbf{x}|$, and $H(r)$ is a function which approaches 1 at $r\rightarrow\infty$ with the asymptotic behaviour~\cite{Forgacs:2005vx}
\begin{equation}
H(r)-1\sim \frac{e^{-m_Hr}}{m_Hr}
\end{equation}
for $m_H<2m_W$.
The Fourier transform of the classical profile $\mathrm{Tr}\,\Phi^2=v^2H(r)^2$ has a delta function peak at $\mathbf{k}=0$, but otherwise it is finite and approaches a constant value at low momenta
\begin{equation}
\langle \mathbf{k}|\mathrm{Tr}\,\hat\Phi^2|\mathbf{0}\rangle\sim \frac{Mv^2}{m_H^3}\quad\mbox{as}\quad k\rightarrow 0.
\end{equation}
For more precise comparison, we will use gradient flow to find the classical monopole configuration ${\Phi,U}$ numerically for our chosen lattice sizes.

Let us then consider the 
magnetic field $\mathbf{B}(\mathbf{x})$. 
This form factor is the most directly measurable quantity characterising magnetic monopoles, because it determines their scattering amplitude with photons.
For a semiclassical monopole, this has
the standard Coulomb form,
\begin{equation}
\mathbf{B} (\mathbf{x}) = \frac{1}{g} \frac{\mathbf{x}}{x^3},
\end{equation}
which has the Fourier transform
\begin{equation}
\label{equ:Coulombff}
\langle \mathbf{k} | \hat{\mathbf{B}} (0) | \mathbf{0} \rangle = i \frac{4 \pi M}{g} \frac{\mathbf{k}}{k^2}
\end{equation}

Note that this result is the same as for a pointlike monopole, which means that the semiclassical calculation is not sensitive to the size or internal structure of the monopole in any way. Therefore it is not useful for probing the properties of magnetic monopoles, and one needs a quantum mechanical result instead.

\subsection{Form factors from two-point functions}

To calculate form factors in quantum theory, we adapt our method given in Ref.~\cite{Rajantie:2010fw} for obtaining the scalar form factor from simpler two-dimensional lattice simulations (and associated one-dimensional defects) to the present case.
Matrix elements like $f$ cannot be computed directly using Monte Carlo simulations. Instead, 
the basic observable is the field correlation function, which we consider in the ground state $|0\rangle$ of the one-monopole sector.
We calculate this correlation function in momentum space, taking the Fourier transform in space but not in the Euclidean time direction, and write a spectral expansion in terms of energy eigenstates $|\alpha\rangle$ with energies $E_\alpha$,
\begin{multline}
\label{equ:spectral}
\langle \mathcal{O}(0,\mathbf{k}) \mathcal{O}(t,\mathbf{q})\rangle \\ =\sum_\alpha
\frac{\langle 0| \hat{\mathcal{O}}(\mathbf{k})|\alpha\rangle\langle\alpha| \hat{\mathcal{O}}(\mathbf{q})|0\rangle}{\langle0|0\rangle}
e^{-t(E_\alpha-E_0)},
\end{multline}
where $E_0=M$ is the energy of the single-monopole ground state.

Furthermore, the Euclidean spacetime is necessarily finite in actual Monte Carlo simulations. We denote the length of the system in the time direction by $T$ and in the three space directions by $L$. We apply twisted boundary conditions (\ref{eq:bctwisted}) to the spatial boundaries. In addition to creating an odd magnetic charge, this has the effect that all observables that are odd under charge conjugation such as $\mathbf{B}$ are antiperiodic, and even observables such as $\mathrm{Tr}\,\Phi^2$ are periodic. Their momenta $\mathbf{k}$ and $\mathbf{q}$ in Eq.~(\ref{equ:spectral}) are therefore also quantised accordingly,
\begin{equation}
k_i=\left\{
\begin{array}{ll}
(2n_i+1)\pi/L, & \text{for odd operators,}\\
2n_i\pi/L, & \text{for even operators,}
\end{array}\right.
\end{equation}
with $n_i\in\mathbb{Z}$.

In the time direction, we impose periodic boundary conditions.
The correlator can then be written as
\begin{multline}
\label{eq:bigtrace}
\left< \mathcal{O} (0,\mathbf{k}) \mathcal{O}(t,\mathbf{q}) \right> = \frac{1}{Z} \text{Tr}\; \hat{U}(T-t) \hat{\mathcal{O}}(\mathbf{q}) \hat{U}(t) \hat{\mathcal{O}}(\mathbf{k})
\\
=\frac{1}{Z}\sum_{\alpha,\alpha'}\langle \alpha'|\hat{\mathcal{O}}(\mathbf{q})|\alpha\rangle
\langle\alpha|\hat{\mathcal{O}}(\mathbf{k})|\alpha'\rangle e^{-E_{\alpha'}(T-t)-E_\alpha t},
\end{multline}
where $\hat{U}(t)=\exp(-\hat{H} t)$ is the Euclidean time evolution operator, and $Z= \mathrm{Tr}\;\hat{U}(T)$.

With twisted boundary conditions, the states $|\alpha\rangle$, $|\alpha'\rangle$ must have odd magnetic charge, and because of momentum conservation, they must also have opposite overall momentum $\mathbf{k}=-\mathbf{q}$. 
The lowest such state is the single-particle state of a monopole with momentum $\mathbf{k}$, which has energy
\begin{equation}
\ene{k}=\sqrt{k^2+M^2}\approx M+\frac{k^2}{2M}.
\end{equation}
The next states in the spectrum are two-particle states with a monopole moving at momentum $\mathbf{k}'$ and a photon with momentum $\mathbf{k}-\mathbf{k}'$. In a box of size $L$, the momentum of the photon is quantised, and therefore there is a large gap $\sim \pi/L\gg k^2/2M$ between the single-particle state and the lowest two-particle state.

\begin{widetext}
At long time separation, we can therefore approximate Eq.~(\ref{eq:bigtrace}) by an integral over single-particle momentum eigenstates $|\mathbf{k}\rangle$, 
\begin{eqnarray}
\label{equ:fullres}
\left< \mathcal{O} (0,\mathbf{k}) \mathcal{O}(t,\mathbf{q}) \right> &=& 
\frac{1}{Z} \int \frac{d^3 k'}{(2\pi)^3 \ene{k'}}\frac{d^3 k''}{(2\pi)^3 \ene{k''}}\langle \mathbf{k}'|\hat{\mathcal{O}}(\mathbf{q})|\mathbf{k}''\rangle\langle\mathbf{k}''| \hat{\mathcal{O}}(\mathbf{k}) | \mathbf{k}'\rangle e^{-\ene{k'}(T-t)-\ene{k''} t}
\nonumber\\
&=&
\frac{1}{Z} (2\pi)^3 \delta^{(3)}(\mathbf{q}+\mathbf{k}) 
 \int  \frac{d^3 k'}{(2\pi)^3}
\frac{|f(\mathbf{k}'-\mathbf{k},\mathbf{k}')|^2}{\ene{k'-k}\ene{k'}}
e^{-\ene{k'}(T-t)-\ene{k'-k}t}.
\end{eqnarray}
\end{widetext}
Similarly,
we can write the partition function as
\begin{eqnarray}
\label{equ:denominator}
Z &=& \int\frac{d^3 k'}{(2\pi)^3 \ene{k'}}\langle \mathbf{k}'| \hat{U}(T)| \mathbf{k}'\rangle=L^3\int\frac{d^3 k'}{(2\pi)^3}e^{-\ene{k'}T}
\nonumber\\
&\approx& L^3 \int \frac{d^3 k}{(2\pi)^3} e^{-\left(M+\frac{k^2}{2M}\right) T} = L^3\left( \frac{M}{2\pi T} \right)^{3/2}e^{-MT}.
\end{eqnarray}
This partition function is the individual contribution to the partition function from each monopole's worldline in Eq.~(\ref{eq:twistpartfun}), and using Eq.~(\ref{equ:Z1}) it
can be written as
\begin{equation}
Z=Z_1e^{-MT}.
\end{equation}

To calculate the integral (\ref{equ:fullres}), we use the saddle point approximation.
The saddle point $\mathbf{k_0}$ is found by minimising the action
\begin{equation}
\label{equ:pointaction}
S(\mathbf{k}') =   \ene{k'} (T-t) + \ene{k'-k} t - MT
\end{equation}
for given $t$. By approximating the integral by a Gaussian around the saddle point, we obtain
\begin{multline}
\langle \mathcal{O}(0,\mathbf{k})\mathcal{O}(t,\mathbf{q})\rangle = 
\frac{1}{Z}(2\pi)^3 \delta^{(3)}(\mathbf{q}+\mathbf{k})\times \\
\int  \frac{d^3 k'}{(2\pi)^3} \frac{|f(\mathbf{k}'-\mathbf{k},\mathbf{k}')|^2}{\ene{k'-k}\ene{k'}} e^{-S(\mathbf{k}_0)-\frac{1}{2}(\mathbf{k}'-\mathbf{k}_0)\cdot\mathbf{M}(\mathbf{k}_0)\cdot(\mathbf{k}'-\mathbf{k}_0)},
\end{multline}
where $\mathbf{M}(\mathbf{k}_0)$ is the Hessian matrix with components
\begin{equation}
M_{ij}(\mathbf{k}_0)=\left.\frac{\partial^2S(\mathbf{k}')}{\partial k'_i\partial k'_j}\right|_{\mathbf{k}'=\mathbf{k}_0}.
\end{equation}
In the limit of large $t$ and $T-t$, the Gaussian approaches a delta function
and we can calculate the integral\footnote{We have corrected a typographical error in Eqs. (17) and (19) of Ref.~\cite{Rajantie:2010fw} where the square root erroneously extends over the energies as well as the normalisation factor in the denominator, but not the action $S(k_0)$. The expressions used in the numerical analysis were correct and so the results of that paper are unaffected.}
\begin{multline}
\langle \mathcal{O}(0,\mathbf{k})\mathcal{O}(t,\mathbf{q})\rangle = \frac{1}{Z_1} (2\pi)^3 \delta^{(3)}(\mathbf{q}+\mathbf{k}) \\
\frac{|f(\mathbf{k}_0 -\mathbf{k},\mathbf{k}_0)|^2}{\ene{k_0-k}\ene{k_0}} \frac{1}{(2\pi)^{3/2} \, W(\mathbf{k}_0)}e^{-S(\mathbf{k}_0)},
\end{multline}
where
\begin{equation}
W(\mathbf{k}_0) = \sqrt{\det\mathbf{M}(\mathbf{k}_0)}.
\end{equation}

In the non-relativistic limit $k\ll M$, where the form factor is a function of the momentum difference only,
we find
\begin{multline}
\label{equ:finalcorr}
\left< \mathcal{O}(0,\mathbf{k}) \mathcal{O}(t,\mathbf{q}) \right>
\\
\approx \frac{(2\pi)^3 \delta^{(3)}(\mathbf{k}+\mathbf{q})}{L^3}  \left(\frac{T}{M}\right)^{3/2} \frac{|f(\mathbf{k})|^2}{\ene{k_0-k} \ene{k_0} W(k_0)} e^{-S(\mathbf{k_0})},
\end{multline}
where we have substituted the expression (\ref{equ:Z1}) for $Z_1$.

We can use Eq.~(\ref{equ:finalcorr}) to determine the form factor from the field correlator. 
For given $\mathbf{k}$ and $t$, we obtain the saddle point $\mathbf{k}_0$ by minimising Eq.~(\ref{equ:pointaction}), and
the form factor is finally given by
\begin{multline}
\label{equ:fresult}
f(\mathbf{k})=\pm i \sqrt{\left< \mathcal{O}(0,\mathbf{k}) \mathcal{O}(t,-\mathbf{k}) \right>} \\ \times \left(\frac{M}{T}\right)^{3/4} \sqrt{\ene{k_0-k} \ene{k_0} W(k_0)} e^{S(\mathbf{k_0})/2}
\end{multline}
for $\mathcal{O}$ odd. The factor of $i$ is not present for even operators, due to parity considerations. 

\subsection{Mass measurements}
\label{sec:massmeas}

The time separation $t$ enters into Eq. (\ref{equ:fresult}) directly -- in parameterising the two-point function -- as well as indirectly, in the saddle-point calculation for $\mathbf{k}_0$. However, since $\mathbf{k} \ll M$  in the current calculation we can take the nonrelativistic limit of the action in Eq. (\ref{equ:pointaction}) and let $\mathbf{k}_0 = \mathbf{k}t/T$ for arbitrary $t$. To order $\mathbf{k}$, there are no $t$-dependent quantities outside of the action in our expression for the form factor. Thus, as expected, at low momenta we effectively have
\begin{multline}
\label{equ:finalcorr2}
\left< \mathcal{O}(0,\mathbf{k}) \mathcal{O}(t,\mathbf{q}) \right> \\ = \frac{|f|^2}{M^2 } e^{-\sqrt{M^2 + \mathbf{k}_0(t)^2}t - \sqrt{M^2 + (\mathbf{k}-\mathbf{k}_0(t))^2}(T-t) + MT}.
\end{multline}

We can use this result to conduct a fit to the correlator, noting that this is merely one contribution to the two-point function; the other most significant contribution (particularly at shorter distances) will be from the lightest particle that the operator $\mathcal{O}$ can create propagating in the bulk. Hence, if we take $\mathcal{O}=B_i$ then this will be the photon, which we will treat as massless; for $\mathcal{O}=\mathrm{Tr}\,\Phi^2$ it will be the bulk scalar particle, which has a mass $m_H = \sqrt{2}|m|$. 

As always, we are assuming in using this calculation that the particles created by the correlation function either interact directly with the monopole, propagate solely in the bulk, or are annihilated by the vacuum.

\section{Results}

\begin{figure}
\includegraphics[scale=0.34,angle=270]{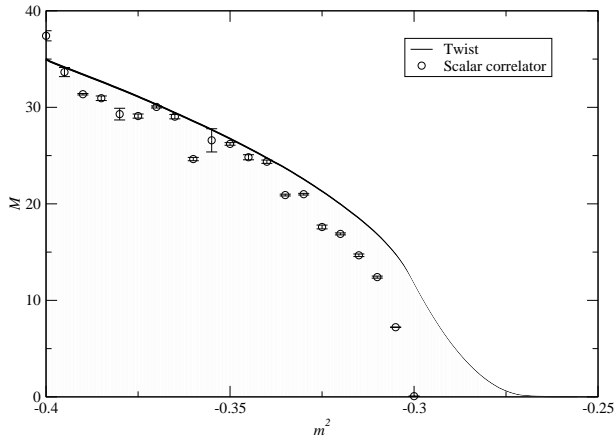}
\caption{\label{fig:monomassscal} Plot of the monopole mass obtained from the scalar field correlator, with twist measurement overplotted for comparison. The scalar field correlator with $\mathbf{k}=(2\pi/L,0,0)$ and permutations is used, the lowest permitted nonzero momentum.}
\end{figure}

\begin{figure}
\includegraphics[scale=0.34,angle=270]{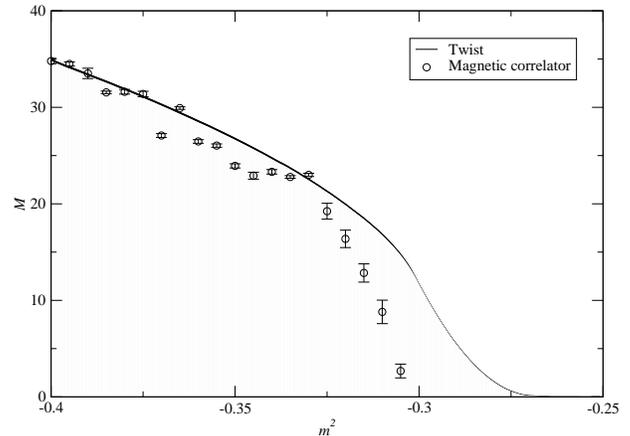}
\caption{\label{fig:monomassbfield} Plot of the monopole mass obtained from the magnetic field correlator, with twist measurement overplotted for comparison. The momenta used are $\mathbf{k}=(3\pi/L,\pi/L,\pi/L)$, and permutations. The results are similar to the scalar field case.}
\end{figure}

Simulations were carried out using a $16^3 \times 48$ lattice with $\lambda=0.1$ and $g=1/\sqrt{5}$. For simplicity, the parameters are the same as those of Ref.~\cite{Rajantie:2005hi}.
For these parameters, the theory has a second-order (or possibly very weakly first-order) phase transition at $m_c^2\approx -0.27$ between the confining phase at $m^2>m_c^2$ and the Coulomb phase at $m^2<m_c^2$.

Configurations were created by generating a classical `cold' monopole and then `heating' the configuration gradually towards the phase transition. Once thermalisation of $\mathrm{Tr}\,\Phi^2$ had occurred for a given parameter choice, the resulting configuration was used as the input for the next value of $m^2$. In this way, a set of configurations was generated which could then be simulated separately. We thermalised the system initially deep in the broken phase, and then gradually increasing the value of $m^2$, because moving through the phase transition in the opposite direction produces extra monopoles which would take a very long time to annihilate~\cite{Kibble:1976sj,Rajantie:2002dw}.

The system seems very susceptible to the creation of metastable states, particularly long-lived monopole-antimonopole pairs as well as what appear to be excited states of the monopole. After thermalisation of $\mathrm{Tr}\, \Phi^2$, additional checks on the histogram of total charge of the system were carried out; fluctuations due to the finite volume (indicated by Eq.~(\ref{eq:twistpartfun})) were to be expected, but any skewness in the distribution led us to reject the thermalisation and try again.

\subsection{Mass measurements and comparison}

Three methods were used to measure the mass of the monopole. The first was the well-established response to a twist obtained from Eq.~(\ref{equ:Fintegral}), used previously in Ref.~\cite{Rajantie:2005hi} and described in detail in Appendix~\ref{app:twist}. The lattices in that work were considerably smaller in the Euclidean time direction but the response to a twist measured here is in good agreement with the $L=16$ data that were obtained. The measurements are plotted as a continuous line in both Figures \ref{fig:monomassscal} and \ref{fig:monomassbfield}, for reasons discussed at length in Section~\ref{sec:twist}. The thickness of the line is the estimated error.

It was found that, for the histograms to offer sufficient overlap that the free energy estimates (\ref{equ:f1}) and (\ref{equ:f2}) agreed within errors, a measurement spacing of at most $\delta m^2 = 0.001$ was required. The solid line plotted therefore required in excess of 150 separate simulations to keep systematic errors at an acceptable level, although Eq.~(\ref{equ:f1}) gives  consistently a lower value than Eq.~(\ref{equ:f2}).

The twist measurements clearly have a finite size effect (also seen in Ref.~\cite{Rajantie:2005hi}) that affects the measurements of the monopole mass when the physical size of the monopole almost fills the box. The curve of the twist results changes concavity as the monopole becomes smaller than the box size. Deeper in the broken phase the monopole mass behaves in a manner similar to the classical monopole mass.

The errors were obtained using the methods described in Appendix~\ref{app:twist}. No attempt was made to account for the nonzero covariance between adjacent mass interval measurements, but it is felt that this would not give a major systematic contribution to the error.

Let us now turn our attention to the use of the two-point correlator to calculate masses as described in Section~\ref{sec:massmeas}. We carried out a fit to Eq.~(\ref{equ:finalcorr2}), plus a bulk field which we expect will be either the scalar or the photon, depending on the operator used:
\begin{multline}
C(t) = 
 C_1 \, e^{-\sqrt{M^2 + \mathbf{k}_0(t)^2}t - \sqrt{M^2 + (\mathbf{k}-\mathbf{k}_0(t))^2}(T-t) + MT} \\
+ C_2 \, \left( e^{-E_\text{bulk} t} + e^{-E_\text{bulk} (T-t)} \right) .
\end{multline}

To ensure that our error estimates are robust despite the clear correlations between data points at different separations exhibited by the two-point function, we use a jackknifed nonlinear least squares fit method. The error estimates obtained from this technique are (in the present work) in agreement with those from our previous use of bootstrapping, but there exist results demonstrating the robustness of the jackknife technique for residuals that are not independently and identically distributed~\cite{Shao}.

The length of each simulation run was about ten times that for each simulation used in the response to a twist technique discussed above. On the other hand, for the single point at $m^2=-0.4$, the results of 150 such simulations in two topological sectors are required for the twist calculation (given the conditions above of a spacing where the two measurements $f_1$ and $f_2$ agree to within $2\sigma$), whereas just one measurement in the topologically nontrivial sector is needed with the correlator calculation. Added to the difficulty of thermalising every one of those 300 ensembles and avoiding metastability, it becomes clear that it is computationally less demanding to use the correlator measurement deep in the broken phase -- if it can be relied upon. Close to the phase transition, the finite size effects of either technique are severe in such a small box. We therefore leave it to future work to study the dynamics of quantum monopoles at strong coupling near the critical point%
~\cite{Rajantie:2005hi}.

Based on Figure~\ref{fig:monomassscal}, it seems that there are some small systematic discrepancies between the twist and correlator results when the correlator of the scalar field is used. Given the relatively small lattice sizes used it is not inconceivable that this is due to the finite size effect in one of the two quantities measured, but long-lived metastable states are another possibility. Note that we do not anticipate any major lattice artefacts playing a role in the monopole dynamics until $m_\mathrm{H} \approx 1$, when the scalar mass is about the inverse lattice spacing, at which time the monopole will become small enough to feel the potential due to the discretised lattice more severely~\cite{Speight:1998uq}.

The results for the magnetic field correlator are shown in Figure.~\ref{fig:monomassbfield}. Since the magnetic field operator couples to the photon, we anticipate that part of the signal in this case comes from a massless photon field propagating in the bulk. This assumption seems borne out by the failure of our fitting ansatz for $\mathbf{k}=(\pi/L,\pi/L,\pi/L)$, and the need to go to $\mathbf{k}=(3\pi/L,\pi/L,\pi/L)$ to see the correlator expected of the monopole signal.

Despite the apparent systematic discrepancy, the fits yielding the data for  Figure.~\ref{fig:monomassbfield} are very good, and the form of the correlator given in Eq.~(\ref{equ:finalcorr2}) seems to be the right one; the long distance `plateau' behaviour is a good fit.

\subsection{Form factor measurements}

Having studied the mass using the low-momentum correlator measurements, we now move on to the form factor measurements. 
From the results of the Monte Carlo simulations we use Eq.~(\ref{equ:fresult})
to obtain the form factor, and compare with semiclassical expectations. To minimise sources of systematic error, we use the twist results for the value of $M$ in computing form factors.

\begin{figure}[t]
\includegraphics[scale=0.34,angle=270]{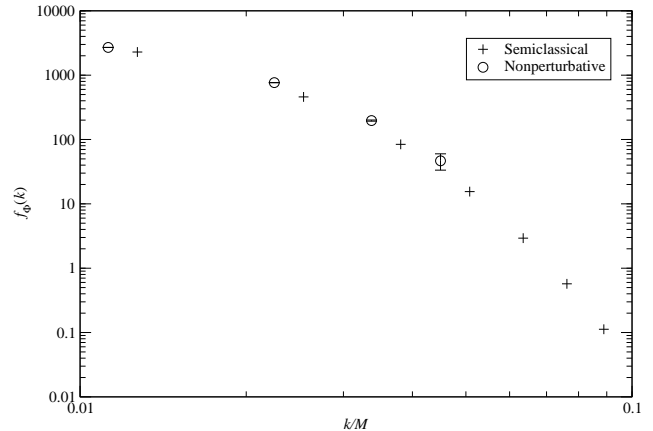}
\caption{\label{fig:monoffscal} Plot of the monopole form factor for the scalar field $f_\Phi(k)$. The measurement is deep in the broken phase with $m^2 = -0.4$. For comparison, the semiclassical result is also shown; renormalisation conditions have been used such that the vacuum expectation value for the gradient flow monopole matches that measured in the nonperturbative simulation; its classical mass is then $M_\text{cl}=30.9$, to be compared with $M=34.9\pm 0.1$ for the quantum monopole.}
\end{figure}

We start by looking at the scalar field form factor $f_\Phi(k)=\langle \mathbf{k}|\mathrm{Tr}\,\hat\Phi^2|\mathbf{0}\rangle$, for which there is a semiclassical comparison available. The classical monopole configuration can be obtained on the lattice using gradient flow. Following Eq.~(\ref{equ:classff}), the form factor can then be recovered by Fourier transforming $\mathrm{Tr}\,\Phi^2$, for comparison with the measurement from the Monte Carlo simulation.

The minimisation was started from a classical `hedgehog' (\ref{equ:hedgehog}) with a trivial gauge field $U_\mu (x) = 1$. A local minimum of the Euclidean action was obtained using gradient flow (see Appendix \ref{app:gradflow} for details).
The resulting field configurations were used to obtain the scalar field $\mathrm{Tr} \, \Phi^2$ in the presence of the classical monopole. As a by-product the classical mass was obtained (for comparison with the twist results above), by looking at the difference in energy between topologically trivial and topologically nontrivial configurations,
\begin{equation}
M_\text{cl} (m^2) = E_\text{tw} (m^2) + \frac{m^4}{4\lambda}L^3.
\end{equation}

Our results deep in the broken phase are shown in Figure~\ref{fig:monoffscal}. In this plot, a single value of $m^2=-0.4$ has been used for the Monte Carlo simulations, and the classical monopole with the closest matching mass was used for the comparison. There is, unsurprisingly, good agreement between the two.
The semiclassical agreement demonstrates that our technique generalises reliably from the relatively straightforward case of the kink to higher dimensions.

\begin{figure}[t]
\includegraphics[scale=0.34,angle=270]{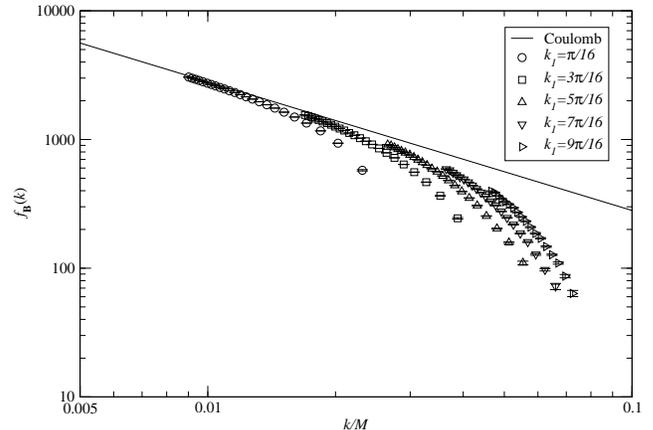}
\caption{\label{fig:monoffbfield} Plot of the monopole form factor for the magnetic field
$f_\mathbf{B}(k)$, for momenta $(k_1, \pi/L, \pi/L)$ and permutations on a doubly logarithmic scale. The comparison is with the Coulomb result.}
\end{figure}

The magnetic field form factor $\mathbf{f}_\mathbf{B}(\mathbf{k})=\langle\mathbf{k}|\mathbf{\hat{B}}|\mathbf{0}\rangle$, is perhaps physically more interesting. It is a vector quantity, but in continuum its direction is always parallel to $\mathbf{k}$ because of rotation invariance, and therefore only its length $f_\mathbf{B}(k)=|\mathbf{f}_\mathbf{B}(\mathbf{k})|$ is non-trivial.
On the other hand, in the simulations it is easiest to consider its individual components $\left[\mathbf{f}_\mathbf{B}(\mathbf{k})\right]_i$, but because of the boundary conditions we cannot choose the momentum to be parallel to a coordinate axis. Instead, we note that
the length of the vector can be written as
\begin{equation}
f_\mathbf{B}(k)=\frac{k}{k_i}\left[\mathbf{f}_\mathbf{B}(\mathbf{k})\right]_i.
\end{equation}
This quantity is shown for various values of $M(m^2)$ and $k$ in Figure~\ref{fig:monoffbfield}. 
For $k\ll m_H$, we are probing wavelengths longer than the monopole core size, and therefore
the curve approaches the expected Coulomb result of Eq.~(\ref{equ:Coulombff}). 
In the semiclassical calculation this behaviour extends to arbitrarily high momenta, which corresponds to a pointlike charge, but our results show that in the quantum theory there is
a clear deviation from the Coulomb result at 
at shorter wavelengths, when $k\gtrsim m_H$. One interpretation for this is that because of quantum fluctuations, the magnetic charge is spread out over distance $\sim 1/m_H$.

\begin{figure}[t]
\includegraphics[scale=0.34,angle=270]{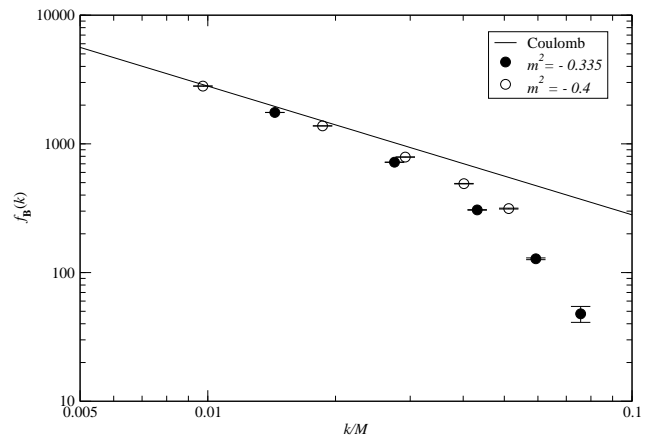}
\caption{\label{fig:monoffbfieldsingle} Plot of the monopole form factor for the magnetic field $f_\mathbf{B}(k)$, for $m^2=-0.335$ and $m^2=-0.4$, which can be interpreted as two different lattice spacings. The Coulomb result is again included for comparison.}
\end{figure}

In Figure~\ref{fig:monoffbfieldsingle}, we highlight two fixed values of $m^2$ and plot the form factor for various values of $k$. Changing the value of $m^2$ can be interpreted as changing the physical lattice spacing. Moving closer to the critical point, i.e., towards higher $m^2$, correspond to taking the continuum limit. In Fig.~\ref{fig:monoffbfieldsingle}, we see that closer to the continuum limit, the charge distribution becomes more spread out in physical units. On physical grounds we would expect that it approaches a finite continuum limit.

\subsection{A note on algorithms and performance}

Previous nonperturbative studies of topological solitons have typically employed a standard Metropolis update algorithm. It would seem, however, that excitations corresponding to an extended defect's  worldline are not going to be quickly thermalised or decorrelated by updates that are local in space. Indeed, the classical topological soliton is a solution of the field equations, so an obvious way to improve ergodicity would seem to be to use one of the family of algorithms which relies on real-time dynamics. For this reason, despite the added computational cost -- and the complexity arising from the twisted boundary conditions -- it was decided to investigate the performance of a Hybrid Monte Carlo (HMC) algorithm. This showed promise previously when fighting critical slowing down in our studies of the form factors of critical kinks. We hypothesised that this was due to the fact that the defects obey the equations of motion, and so using an update method that integrates the equations of motion (or a generalisation thereof) improves ergodicity for observables associated with the quantum topological soliton. In contrast, a single Metropolis update step will not significantly alter the position or configuration of a topological defect.

In the current situation, however, it was difficult to detect an advantage to using HMC. While the autocorrelation time was in many cases the same, the CPU time required to integrate a single trajectory was longer than a single Metropolis checkerboard sweep; the staples must be recalculated for every step in the trajectory. This poor performance may be due to our being at relatively weak coupling, with severe finite size effects that mask any critical slowing down. It may also be due to inadequate tuning of the HMC algorithm to give an optimal acceptance rate.

Lastly, to improve statistics for the magnetic field correlator we considered an overrelaxation step for the $\mathrm{SU}(2)$ gauge fields, coupled to an accept-reject step to account for the covariant derivative term in the action. We made use of the $\mathrm{SU}(2)$ move $U\to U_0 U^{-1} U_0$. This leaves the Wilson term unchanged when $U_0 = V^{-1} \sqrt{\mathrm{det}\; V}$, where $V$ is the `staple'~\cite{Creutz:1987xi}. Unfortunately we did not notice any substantial improvement to the statistics as a result of adding this step.

\section{Conclusions}

We have used correlation functions to measure properties of the 't Hooft-Polyakov monopole nonperturbatively.
For the monopole mass we found good agreement with previous studies that used the response to twisted boundary conditions.

We also calculated the form factors of the monopole for scalars and photons. The form factor for the photon is physically more relevant, because it describes the interaction of the monopole with photons and the pair creation of monopoles through the Drell-Yan process.
It is, therefore, the most relevant quantity for accelerator experiments~\cite{Fairbairn:2006gg}. It has been argued~\cite{Drukier:1981fq} that for 't~Hooft-Polyakov monopoles it is suppressed relative to pointlike Dirac monopoles by many orders of magnitude. However, in order to calculate the pair creation rate from our results we would have to analytically continue the form factor to imaginary momenta, which is not straightforward. 
On the other hand, if monopoles are produced at the LHC, then one can envisage further experiments that probe their properties in much more detail such as scattering involving other particles. The form factor for real momenta, which we have calculated, is directly relevant for such processes.

For the scalar we find good agreement with the semiclassical results, which was expected because of the weak coupling. The semiclassical form factor for the photon is that of a pointlike magnetic charge, but our results indicate a smooth charge distribution in the full quantum theory. This shows that a proper quantum calculation is absolutely necessary in order to probe the internal structure of monopoles using photons. The continuum limit deserves to be explored using the same techniques.

It should be reiterated that although our expression for the magnetic field -- Eq.~(\ref{eq:latmag}) -- has attractive properties, it is not the exact creation operator for asymptotic photon states in the full quantum theory. In principle, a numerical approximation for the correct creation operator could be obtained by a diagonalisation procedure.

In Ref.~\cite{Edwards:2009bw}, it was shown how to generalise the twisted boundary conditions to other $\mathrm{SU}(N)$+Higgs models, $N$ even. Although odd $N$ is arguably of greater phenomenological interest, the techniques demonstrated here should be equally valid in these cases.

\acknowledgments

This work was supported by the Science and Technology Facilities Council and the Royal Society. We have made use of the Imperial College High Performance Computing Service. Some calculations for this paper were performed on the COSMOS Consortium supercomputer within the DiRAC Facility jointly funded by STFC and the Large Facilities Capital Fund of BIS.

\appendix
\section{Twist measurement}
\label{app:twist}
To determine the monopole mass from the free energy difference, as discussed in Section~\ref{sec:twist}, one needs to integrate its derivative along a path in the parameter space~\cite{Ciria:1993yx,Kajantie:1998zn,deForcrand:2000fi,Rajantie:2005hi}.
The free energy of an ensemble is defined as $F=-\ln Z$, where 
\begin{equation}
Z=\int{\cal D}U_i{\cal D}\Phi e^{-S_\text{lat}}
\end{equation} 
is the partition function.
The derivative of the free energy difference~(\ref{equ:deltaF}) is therefore
\begin{equation}
\frac{\partial \Delta F}{\partial g} = \left[ \left\langle \frac{\partial S_\text{lat}}{\partial g}\right\rangle_\mathrm{tw}  - \left\langle\frac{\partial S_\text{lat}}{\partial g}\right\rangle_\mathrm{C} \right],
\end{equation}
where the subscripts indicate expectation values calculated in the two ensembles.
In our calculations, we take $g=m^2$, yielding
\begin{equation}
\label{eq:massderiv}
\frac{\partial \Delta F}{\partial m^2} = \left[ \left\langle \mathrm{Tr}\,\Phi^2\right\rangle_\mathrm{tw}  - \left\langle \mathrm{Tr}\,\Phi^2\right\rangle_\mathrm{C} \right].
\end{equation}

Integrating this expression from the symmetric phase where $M=0$ through the phase transition to our desired value of $m^2$ will yield, in principle, yield the mass $M$. However, it is difficult to obtain reliable error estimates from this technique; it is important that we keep the error estimates under control.

In practice, one uses finite differences instead of the derivative (\ref{eq:massderiv}).
The free energy difference between two different values of $m^2$ can be written in two ways,
\begin{equation}
\label{equ:f1}
f_1=-\ln\left\langle e^{-(m^2_2-m^2_1)\sum_x {\rm Tr}\,\Phi^2}\right\rangle_1
\end{equation}
and 
\begin{equation}
\label{equ:f2}
f_2=\ln\left\langle e^{-(m^2_1-m^2_2)\sum_x {\rm Tr}\,\Phi^2}\right\rangle_2
\end{equation}
where the expectation values are calculated at $m_1^2$ and $m_2^2$.
Having established that the two measurements are in agreement, the change in the monopole free energy is
\begin{equation}
\label{eq:masschange}
\Delta F(m_2^2) - \Delta F(m_1^2) = \frac{1}{2}\left(f_1^\mathrm{tw} + f_2^\mathrm{tw} - f_1^\mathrm{cl} - f_2^\mathrm{cl}\right),
\end{equation}
where we have chosen to average $f_1$ and $f_2$ with equal weights.
The errors for $f_1$ and $f_2$ for each sector are added in quadrature and therefore we have
\begin{multline}
\Delta \left[ \Delta F(m_2^2) - \Delta F(m_1^2) \right]^2 = \\ 
\frac{1}{4} \big[ \Delta f^2_{1,\mathrm{tw}} + \Delta f^2_{2,\mathrm{tw}}  + (f_{1,\mathrm{tw}} - f_{2,\mathrm{tw}})^2 \\ 
+  \Delta f^2_{1,\mathrm{C}} + \Delta f^2_{2,\mathrm{C}}  + (f_{1,\mathrm{C}} - f_{2,\mathrm{C}})^2  \big]
\end{multline}
We know $M=0$ in the symmetric phase, so we start summing the differences from a value of $m^2$ where the symmetry is not yet broken. Whereas the errors in each $f$ for a change from $m_1^2$ to $m_2^2$ are themselves independent, there is a small nonzero covariance for two different adjacent steps. Some care is therefore needed when summing all the errors in a mass measurement.

As a check, we should make sure that our measurements of $f_1$ and $f_2$ are concordant, since they measure the same thing. This ensures that the two ensembles at $m_1^2$ and $m_2^2$ are thermalised and at equilibrium, and the spacing is sufficiently small that the histograms of data for $\mathrm{Tr}\,\Phi^2$ overlap adequately. Ferrenberg and Swendsen's work encourages us to see this process as the reweighting of a histogram of measurements, and their formula immediately yields
\begin{equation}
P_{m_2^2}(\mathrm{Tr}\,\Phi^2) = \frac{P_{m_1^2}(\mathrm{Tr}\,\Phi^2) e^{(m_2^2 - m_1^2)\mathrm{Tr}\,\Phi^2}}{\sum_{\mathrm{Tr}\,\Phi^2} P_{m_1^2}(\mathrm{Tr}\,\Phi^2) e^{(m_2^2 - m_1^2)\mathrm{Tr}\,\Phi^2}}
\end{equation}
for the observed distribution of $\mathrm{Tr} \, \Phi^2$ sampled at $m_1^2$ and evaluated at $m_2^2$~\cite{Ferrenberg:1988yz}; interchanging $m_1^2$ and $m_2^2$ gives the expression for measurements sampled at $m_2^2$ evaluated at $m_1^2$. The equality $f_1=f_2$ then follows, but the importance of this approach is the realisation that the measurements will not agree unless sufficient overlap of the histograms for $\mathrm{Tr}\,\Phi^2$ at both $m_1^2$ and $m_2^2$ are available. This overlap means that the difference in the actions, $\Delta S = \sum_x (m_2^2 - m_1^2)\mathrm{Tr}\,\Phi^2$, should be relatively small.

Less overlap means the inferred value of $\mathrm{Tr}\,\Phi^2$ is an over (or under) estimate, being closer to the original value than required. This will make $f_1$ and $f_2$ too big, so the free energy will be overestimated. A similar overestimate will occur in both topological sectors, although $\sum_x\mathrm{Tr}\,\Phi^2$ is smaller in the topologically nontrivial sector.

Because we can use this approach to accurately interpolate the free energy of the monopole at any value of $m^2$, the result is plotted continuously on Figures~\ref{fig:monomassscal} and \ref{fig:monomassbfield}. If we encounter difficulty obtaining good statistics, then this technique would be ideal to use alongside parallel tempering (replica exchange Monte Carlo), due to the necessarily small separations between values of $m^2$~\cite{SwendsenRMC}. In our case replica exchange would have been usable if the measurement spacings were slightly smaller, but would have led to substantial wait times with our computer cluster.

\section{Gradient Flow}
\label{app:gradflow}
The equations we used to minimise the classical action were
\begin{widetext}
\begin{equation}
\phi^a(x,\tau+\delta \tau) =   \phi^a(x,\tau)  + \delta \tau \bigg[ -\left[4(8+m^2) + 8 \, \lambda \, \mathrm{Tr}\, \Phi^2\right]\phi^a(x,t)   + \sum_j \left[ \sigma^a U_j(x,t) \Phi(x+\hat{\jmath},t) U^\dag_j(x,t)\right] \bigg]
\end{equation}
and
\begin{equation}
U_i(x,\tau + \delta \tau)  = \exp \left\{ i\,\delta t\, \sigma^a  \bigg[ -\frac{\beta}{2}\sum_{\text{staple}} \, \mathrm{Tr} \left\{ \sigma^a U_{ij} (x,\tau) \right\}  + 2 \, \mathrm{Tr} \, \Big\{ \sigma^a U_i(x,\tau) \Phi(x,\tau) U_i^\dag (x,\tau) \Phi(x + \hat{\imath},\tau) - \text{h.c.} \Big\}   \bigg] \right\} U_i(x,\tau) .
\end{equation}
\end{widetext}
The length $T$ of the Euclidean time direction is not relevant for this process and so we could set $T=1$.

\bibliography{monopole}

\end{document}